\documentclass{article}
\usepackage{spconf,amsmath,epsfig,bm}
\usepackage{tabulary}
\usepackage{graphicx}
\usepackage{fixmath}
\usepackage{amssymb}
\usepackage{textcomp}
\usepackage{subfig}
\usepackage{gensymb}

\title{Coding schemes and applications for weather radars}

\makeatletter
\def\@name{ \emph{Mohit Kumar$^1$, V. Chandrasekar$^1$, Shashank Joshil$^1$} \\
            \thanks{This work is sponsored by the NASA GPM project.} }
\makeatother
\address{$^1$ Colorado State University, Fort Collins, CO}

\begin{document}

\maketitle

\section{Introduction}

In this paper, we describe the evolution of a pair of polyphase coded waveform for use in second trip suppression in a weather radar. The polyphase codes were designed and tested on NASA weather radar.
The NASA dual frequency, dual polarization Doppler radar (D3R) was developed primarily as a ground validation tool for the GPM satellite dual frequency radar \cite{Vega2014}. Recently, the D3R radar was upgraded with new versions of digital receiver hardware and firmware, which supports larger filter lengths and multiple phase coded waveforms, and also newer IF sub-systems (see \cite{8128188}, \cite{8517944} and \cite{MohitAms}) . This has enhanced the capabilities of radar manifolds. \par

\begin{figure}
	\centering
	\noindent\includegraphics[width=3in,angle=0]{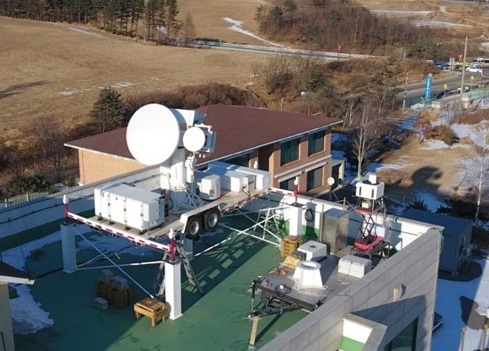}
	\caption{ The NASA D3R deployed at rooftop of DaeGwallyeong Regional Weather Office (DGW) in the daegwallyeong-myeon province of South Korea. }
	\label{fig:radar}
\end{figure}

D3R has taken part in various field campaigns in North America and abroad (see \cite{8127562}, \cite{8517313} and \cite{MohitIcepop}). D3R deployed in the winter Olympic and Paralympic games of 2018 is shown in Figure \ref{fig:radar}.In this paper, we show the performance of intra-pulse coding schemes to mitigate second trip effects. With these new waveforms, better retrievals of polarimetric variables can be obtained.\par
\begin{figure}[!t]
	\centering
	\includegraphics[width=3in]{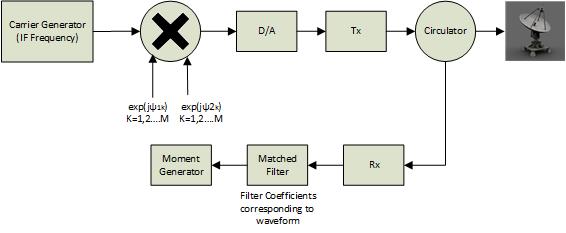}
	\caption{The system architecture to make use of Intra-Pulse coding schemes for second trip suppression.}
	\label{fig:blockDia}
\end{figure}

The intra-pulse (within the pulse) orthogonal polyphase coding, between adjacent pulses (introduced in \cite{8952671}), with minimal ISL (integrated side-lobe level) filters, can effectively suppress second trip echoes from beyond the unambiguous range and minimize its effect on first trip and/or minimize creating non-existential artifacts in the dual-pol moment images.
These intra-pulse codes can be binary or polyphase codes but the mainlobe to sidelobe ratio is better for the polyphase codes of the same length, so polyphase codes are a better choice. And we have used them for our design. Compared to binary codes, the polyphase intra-pulse codes also have better doppler tolerance. However, good auto- and cross-correlation properties are of prime importance in the design of intra-pulse coded waveforms for pulse compression weather radars \cite{7093191}, \cite{381922}. For example, if we design two polyphase sequences which are orthogonal (the two sequences are uncorrelated) and the first and second trip echoes are coded with these sequence set. Then, the alternate pulses will correlate but any two of the adjacent pulses will not correlate. The cross-correlation function between the sequences will give the separation between first and second trip echoes. This can be also explained with Figure \ref{fig:blockDia}. 

Good auto-correlation implies that the synthesized sequence is nearly un-correlated with the time shifted versions of itself and good cross-correlation signify that the sequences are uncorrelated with time shifted versions of any other sequences in the synthesized group of sequences \cite{7093191}. Also, good autocorrelation ensures that the matched/mis-matched filter is able to detect the signal scattered from the range of interest and simultaneously removing the cross talk from other trips or from other range bins. Since the scatterers are continuous in detecting weather phenomenon, the cross talk due to other range bins, is a major concern.\par

This concern is addressed in mean square sense, by using Mismatched filtering techniques \cite{Bharadwaj2012}, which known for reducing range sidelobes and is specially helpful in systems with smaller pulse widths and larger filter size available for pulse compression. The larger is the ratio of the length of the filter to the pulse width, the better chance of the sidelobe energy being spread out, bringing down the peak sidelobe level. This same fact can also be utilized to bring down the cross-correlation sidelobe peaks in the cross-correlation function of orthogonal waveforms. In such a case, the energy function to be minimized, should take both the cross-correlation sidelobe energy and the auto-correlation sidelobe energy into consideration.\par
Thus, the successful design of the orthogonal Intra Pulse codes depends on the degree to which the sequences in the orthogonal set  $\{s_{1}(t),s_{2}(t)\}$ satisfy the property:
\begin{equation} \label{eq_2}
\int_{t} s_{1}(t)s_{2}^{*}(t+\tau)dt = 0, \tau = 1,2,...,M
\end{equation}

\begin{equation}
\int_{t} s_{l}(t)s_{l}^{*}(t+\tau)dt=\begin{cases}
1, & \text{if $\tau = 0, l\in {1,2}$}.\\
0, & \text{otherwise}.
\end{cases}
\end{equation}

which explains concisely the concepts developed earlier. \par

Such polyphase codes are very difficult to design and only optimal approximations can be achieved with pseudo-orthogonality. The cost function, to be minimized, is the total energy in the sidelobes, better known as the integrated sidelobe level (ISL).

\section{Mis-matched Filter Based Orthogonal Polyphase Code Design Problem} \label{section_2}
The motivation behind use of mismatched filter based method, is to have relatively larger filter lengths compared to the sequence length transmitted, so that the peak sidelobe energy can be spread into much larger coefficient space. This approach has more flexibility as we can optimize the sequences and filters separately in an optimization framework, to obtain better sequences.\par
If the polyphase code-filter pairs to be optimized for both auto- and cross-correlation (using mismatched filter), then the error function is given by:
\begin{equation} \label{eq_12}
\varepsilon_{hi} = \textbf{h}_{i}^{t}\varSigma_{i=1}^{k}(\textbf{X}_{i}^{m}\textbf{X}_{i}^{mH})\textbf{h}_{i}^{*}
\end{equation}
where $k$ is the number of codes in the orthogonal set and the optimum filter weights would be:
\begin{equation} \label{eq_13}
\textbf{h}_{i}^{t} = (\varSigma_{i=1}^{k}(\textbf{X}_{i}^{m}\textbf{X}_{i}^{mH})^{-1}\textbf{x}_{i}L / \textbf{x}_{i}^{H}(\varSigma_{i=1}^{k}(\textbf{X}_{i}^{m}\textbf{X}_{i}^{mH}))^{-1}\textbf{x}_{i}
\end{equation}
where $\textbf{h}_{i}$ is the mismatched filter designed for the $i^{th}$ sequence in the orthogonal code set and $\textbf{X}_{i}^{m}$ is the modified transmit convolution matrix for the $i^{th}$ sequence, obtained by deleting the columns of transmit convolution matrix that corresponds to the mainlobe samples (after convolution operation). \par
Our method starts with a set of two codes, randomly generated. The filters based on equation \eqref{eq_13}  were generated (optimal in ISL of the auto- and cross- correlation sidelobes). After we have obtained a set of initial codes and filters, then we iterate over the permissible code combinations and simultaneously trying to find minima for the error function in equation \eqref{eq_13} using optimization methods. In the next iteration, when a new code is found, again we get ISL filter using equation \eqref{eq_13} and the error function is evaluated. This continues until we reach at least, a local minima. The mainlobe width is assumed to be 5 samples. This method jointly optimizes the polyphase codes and filters. Subsequently, global optimization forces the local solvers to run from various initial points in the search space, which are obtained using scatter search.

\subsection{Constraints} \label{sec_cons}
We already mentioned about the uni-modular constraint on the optimization problem. This is important for the transmit section, which might need to operate in saturation mode. Moreover with use of mismatched filter, an overall constraint on the gain and phase terms of the combined code-filter pair, has to be incorporated. Apart from this, there is also a need of another constrain that the gain and phase terms of code-filter pair 1 be equal to gain and phase terms of code-filter pair 2. This would ensure that for the velocity retrieval, the phase imbalance of odd and even pulses, do not give rise to additional spurious velocity components. The gain balance between these pairs, would aid in reflectivity calibration.
\subsection{Ambiguity Function for mismatched filters}
The ambiguity function for a mismatched filter represents the effect of delay and doppler on sidelobes and mainlobe, on a code-filter pair. In general, ambiguity functions are good tools to analyze the performance of synthesized polyphase codes for doppler tolerance in a weather sensing application. In any case, the ambiguity function at $(\tau,f_{d}) = (0,0)$ corresponds to a matched output to the signal reflected perfectly from the target of interest. The ambiguity function $|\chi(\tau,f_{d})|$, for a polyphase code $\textbf{a}$ and mismatched filter $\textbf{b}$, can be written as:
\begin{equation}
|\chi(\tau,f_{d})| = |\varSigma_{n=-N+1}^{N-1} a(n)b(n+\tau) exp(j2\pi f_{d}n)|
\end{equation}

\begin{figure}[!t]
	\centering
	\includegraphics[width=2.5in]{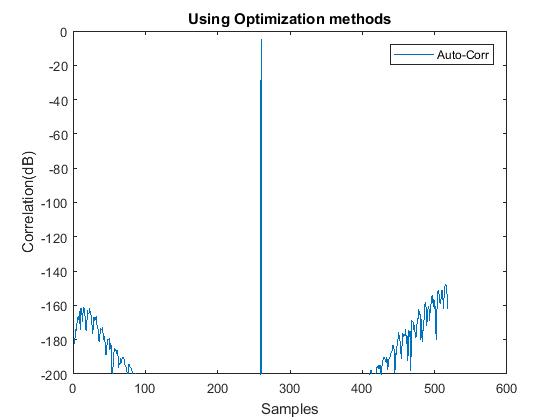}
	\caption{The zero doppler cut of the synthesized polyphase code of length 40 with uni-modular constraint (from auto-ambiguity function).}
	\label{fig_synthcode}
\end{figure}

\begin{figure}[!t]
	\centering
	\includegraphics[width=2.5in]{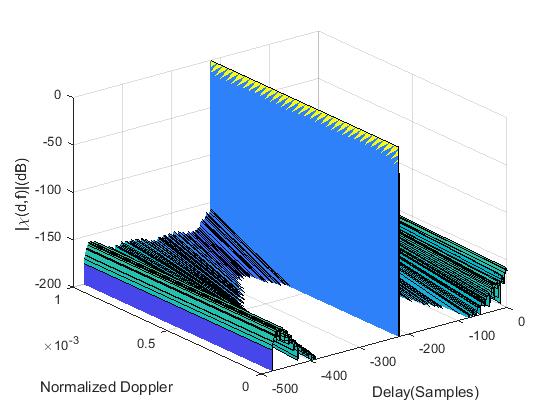}
	\caption{Auto-ambiguity Function of polyphase code with length 40 samples (PW = 20us, BW = 2MHz) and the mismatched filter length of 480 samples.}
	\label{fig_ambcode}
\end{figure}

\begin{figure}[!t]
	\centering
	\includegraphics[width=2.5in]{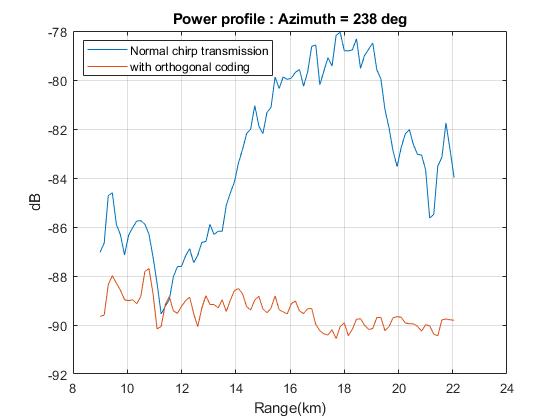}
	\caption{Power profile along ray at radial $238 \degree$ azimuth, clearly depicts the second trip suppression capability of the orthogonal polyphase codes.}
	\label{fig_powerProfile}
\end{figure}

\begin{figure}[!t]
	\centering
	\includegraphics[width=2.5in]{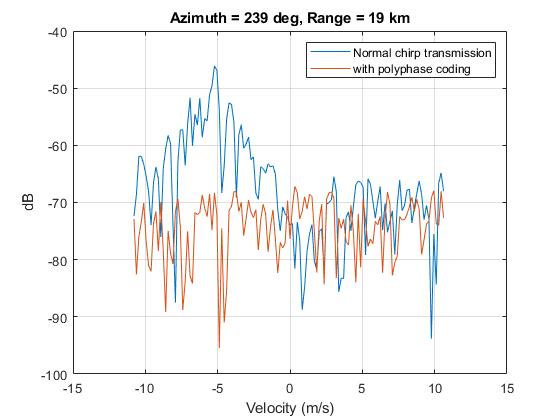}
	\caption{Velocity profile of second trip, observed for normal and polyphase coded transmission. These profiles are for cases shown in (b) and (d) of figure \ref{fig_case1}.}
	\label{fig_velPolt}
\end{figure}
Figure \ref{fig_synthcode} and \ref{fig_ambcode} show the zero-doppler cut (from the ambiguity function) of one pair of synthesized polyphase code and filter with uni-modular constraint. It used minimum sidelobe energy as the error/cost function and mis-matched filter length of 480 coefficients. The zero doppler cut has very low sidelobes in the whole domain, with a one sample mainlobe. The doppler performance of the code is also good under reasonable doppler assumption (peak sidelobe level below -150dB).\par

\begin{figure*}[!t]
	\centering
	\begin{tabular}{c c c}
		\subfloat[]{\includegraphics[width=1.5in]{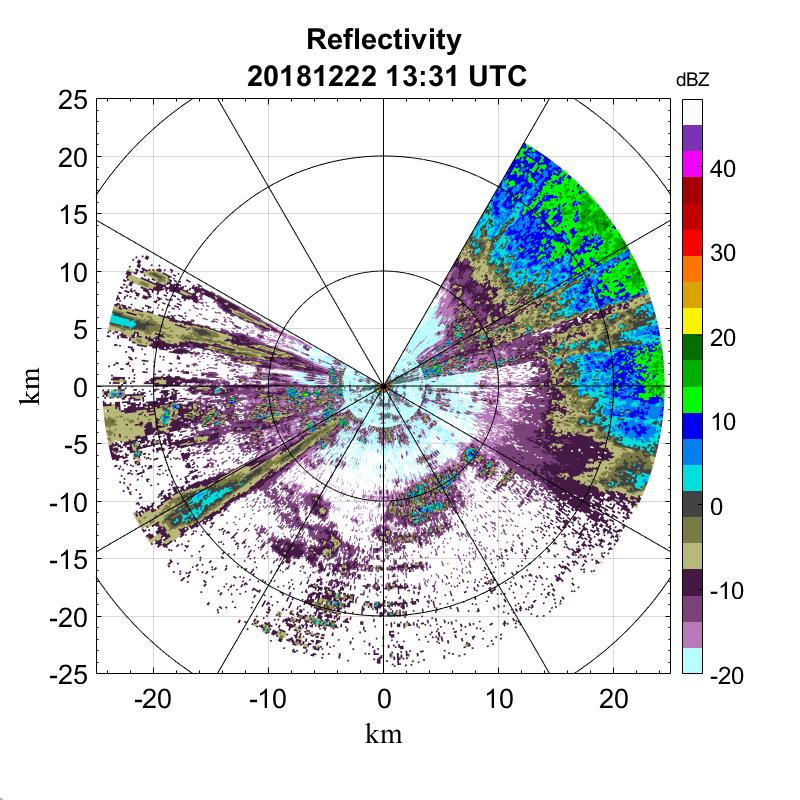}%
			\label{fig_first_case}}
		&
		\subfloat[]{\includegraphics[width=1.5in]{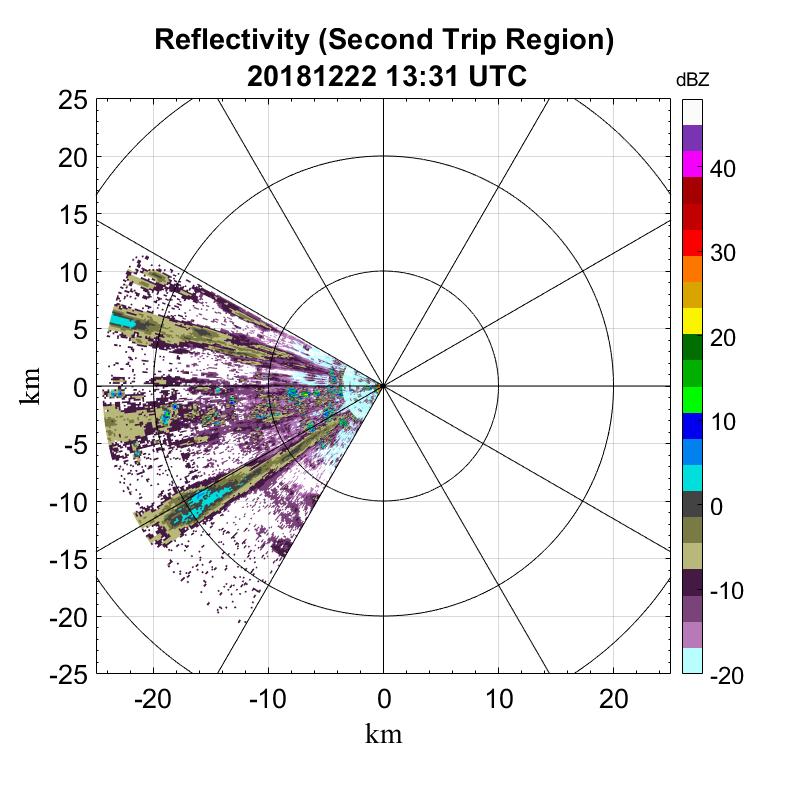}%
			\label{fig_second_case}}
		&
		\subfloat[]{\includegraphics[width=1.5in]{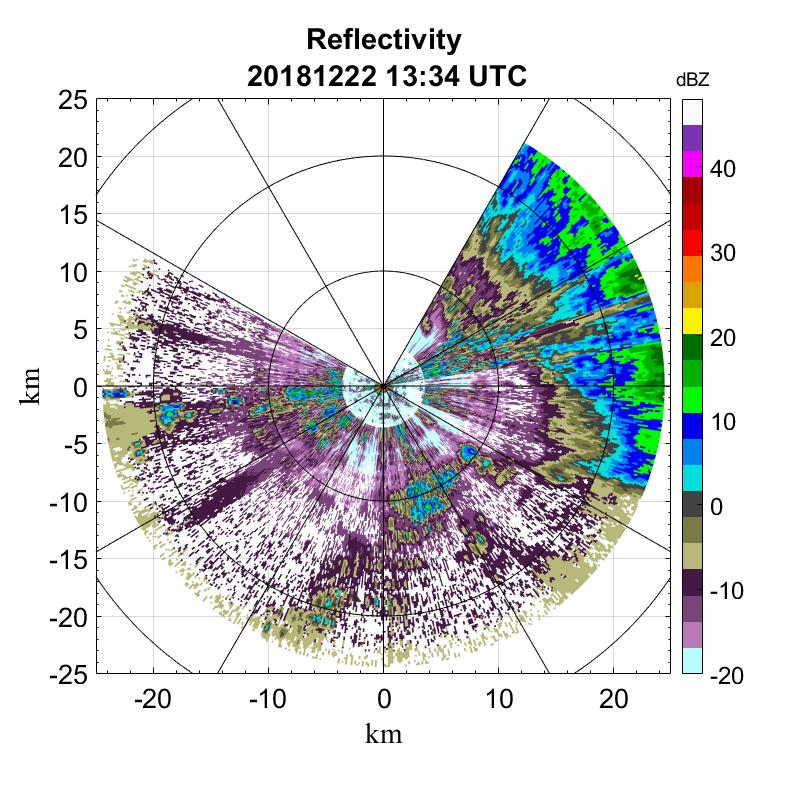}%
			\label{fig_third_case}}
		\\
		
		\subfloat[]{\includegraphics[width=1.5in]{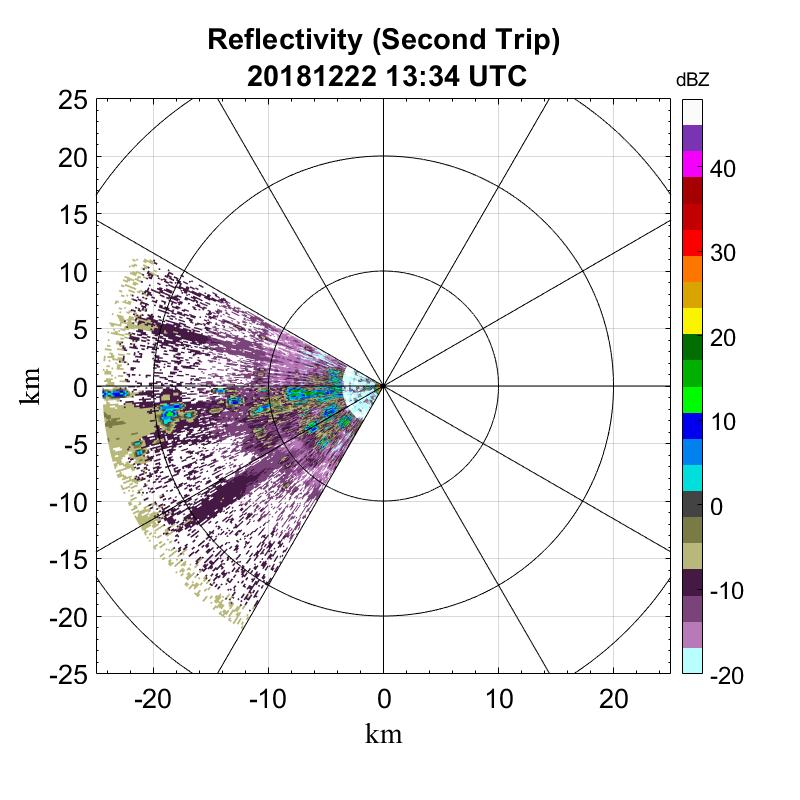}%
			\label{fig_fourth_case}}
		&
		\subfloat[]{\includegraphics[width=2in]{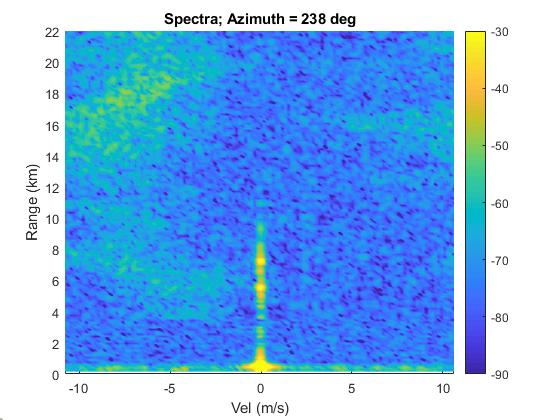}%
			\label{fig_fifth_case}}
		&
		\subfloat[]{\includegraphics[width=2in]{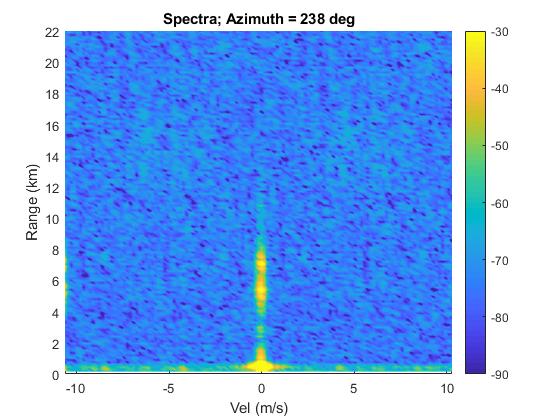}%
			\label{fig_sixth_case}}
		
	\end{tabular}
	\caption{(a) and (b) depict the reflectivity without polyphase codes. It is using a chirp waveform. The south-west region has second trips as confirmed with a Nexrad radar. The first trip lies towards the north-eastern region in this case. (c) and (d) are coded with orthogonal polyphase. The elevation is 2 deg and clearly suppression can be observed at $\sim$ radial 238 $\degree$ azimuth. (e) shows the doppler spectra along the same radial, for normal transmission, whereas, (f) has the same measurement but polyphase code and filter pairs are used.}
	\label{fig_case1}
\end{figure*}

\section{Observations from NASA D3R Weather radar} \label{section_4}
Second trip for D3R would correspond to echoes beyond 75 kms for a $500\mu s$ pri. One of the cases where suppression of second trips with the synthesized polyphase codes could be observed, is depicted in figure \ref{fig_case1}. A good amount of rejection of second trip echoes can be observed in this case, in reflectivity and velocity spectrum, for the polyphase combination. To quantify this, we plot the power profile for the normal transmission and with polyphase coding for this case, along radial at $238 \degree$ azimuth . This is shown in figure \ref{fig_powerProfile}. It can be easily seen that the coding has cleared up the second trip echo and beyond 18 kms of range, it is reduced below noise floor. The velocity plot in figure \ref{fig_velPolt} compares the suppression that the polyphase codes have over normal transmission, in the second trip velocity domain. More than 20dB of reduction in second trip power can be observed from these velocity plots.

\section{Conclusion}
The developed polyphase coded waveforms would enable better retrieval of co-polar moments by suppressing second trip contamination. In this paper, we have shown the development and implementation of intra-pulse polyphase coding for second trip suppression, in D3R radar. More than 20dB of second trip suppression could be observed from reflectivity and velocity profiles. It can be improved further with lower cross-correlation sidelobes, and is a topic of further research.


\bibliographystyle{IEEEbib}
\bibliography{refs1}

\end{document}